\documentclass[aps,twocolumn,groupedaddress]{revtex4}
\usepackage{amsmath}
\usepackage{graphicx}
\begin{document}
\bibliographystyle{apsrev}

\title{Cotunneling Spectroscopy in Few-Electron Quantum Dots}

\author{D. M. Zumb\"uhl}
\affiliation{Department of Physics, Harvard University, Cambridge, Massachusetts 02138}
\author{C. M. Marcus}
\affiliation{Department of Physics, Harvard University, Cambridge, Massachusetts 02138}
\author{M. P. Hanson}
\affiliation{Materials Department, University of California, Santa Barbara, California
93106}
\author{A. C. Gossard}
\affiliation{Materials Department, University of California, Santa Barbara, California
93106}
\begin{abstract} Few-electron quantum dots are investigated in the regime of strong
tunneling to the leads. Inelastic cotunneling is used to measure the two-electron singlet-triplet
splitting above and below a magnetic field driven singlet-triplet transition.
Evidence for a non-equilibrium two-electron singlet-triplet Kondo effect is presented.
Cotunneling allows orbital correlations and parameters characterizing
entanglement of the two-electron singlet ground state to be extracted from dc transport.
\end{abstract}

\maketitle

Transport studies of few-electron quantum dots  have proven to be a rich
laboratory for investigating the energetics of electrons in artificial
atoms \cite{Tarucha, Kouwenhoven, Laterals} as well as related spin
effects, including ground-state spin transitions \cite{OtherFewElectrons,
Sachrajda, Fujisawa, Golovach04}, spin lifetimes \cite{Golovach04,
Fujisawa, Elzerman} and Kondo effects \cite{Kondo, Sasaki, STKondo}. The
interplay of electron-electron interactions, electron spin, and coupling
to a Fermi sea makes transport in the few-electron regime a subtle problem
in many-body physics
\cite{Golovach,STCrossTh,STKondoEx,STKondoInclZeeman,STKondoSingleMode,
Reimann}. Of particular importance is the two-electron case (``quantum dot
helium'') \cite{Reimann} since this is a paradigm for the preparation of
entangled electronic states \cite{Burkard}, and in double quantum dots is
the basis of a quantum gate proposal \cite{LossDiVincenzo}.

In this Letter, we present a detailed experimental investigation of
cotunneling through quantum dots containing one, two, and three electrons. Measurements
of inelastic cotunneling are used to extract the singlet-triplet (ST)
splitting across the two-electron ST transition. Evidence of a
non-equilibrium ST Kondo effect for two electrons is presented.
Cotunneling and Kondo effects are used to determine the g-factor for
magnetic fields along different directions in the plane of the 2D electron
gas (2DEG), giving isotropic g-factors close to the bulk GaAs value. Using
both cotunneling and sequential tunneling data, we extract quantum
correlations of the two-electron singlet ground state, allowing the degree
of spatially separated entanglement to be measured.

Previous transport studies of few-electron quantum dots have identified
the ST ground state transition for two electrons \cite{Tarucha,
Kouwenhoven, Fujisawa, Sachrajda, OtherFewElectrons} as well as for larger
electron numbers \cite{Kogan, STManyE}. Inelastic cotunneling was recently
investigated in few-electron vertical structures in Ref.\,\cite{Silvano}.
These authors demonstrated that inelastic cotunneling provides a direct
and sensitive measure of excited state energies. Here, we use this fact to
measure the ST splitting, $J$, across the ST transition (for both negative
and positive $J$), and for the first time extract two-electron ground
state wave function correlations from cotunneling.

\begin{figure}[b]
             \label{fig1}
\vspace{-0.1in}
\includegraphics[width=3.2in]{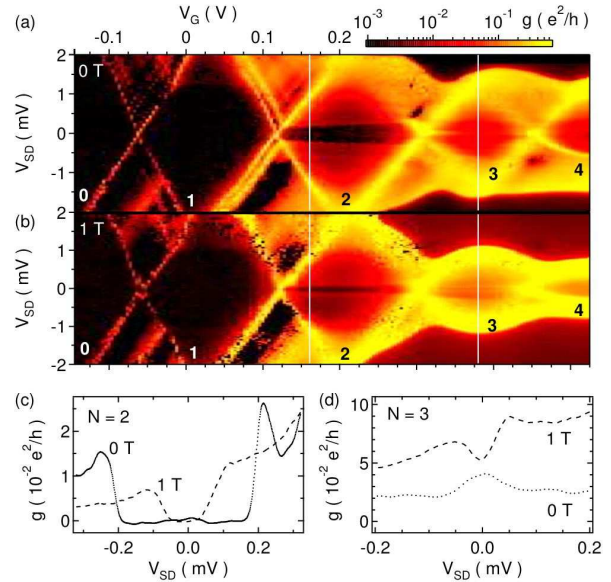}
       \caption{\footnotesize {(a) Differential conductance $g$ (log color
       scale) as a function of source-drain bias $V_{SD}$ and gate voltage
       $V_G$ at $B_\perp=0$, at base electron temperature $T_{el}=45$~mK.
       Numbers 0 through 4 are  number of electrons in the dot.
       White vertical lines identify the locations for data shown in (c) and (d).
       (b) Same as (a), at $B_\perp=1$~T.
       (c) Differential conductance through the $N=2$ diamond showing
       step with overshoot at $V_{SD} = J(B_\bot)/e$ at $B_\bot=0$ and $1$~T.
       (d) Differential conductance through the $N=3$ diamond showing
       Kondo peak at  $V_{SD} = 0$ for $B=0$, split by $B_\perp=1$~T.}}
\end{figure}

Transport through the ST transition has been studied theoretically
\cite{STCrossTh}, with a prediction of enhanced Kondo correlations at the
ST crossing \cite{STKondoEx}. Effects of lifting spin degeneracy of the
triplet have also been theoretically investigated
\cite{STKondoInclZeeman}. For the degenerate triplet case, a
characteristic asymmetric peak in conductance at the ST crossing has been
predicted \cite{STKondoSingleMode, Golovach}. This predicted asymmetric
peak is observed in the present experiment. Previous measurements of ST
Kondo effects \cite{Sasaki, STKondo, tubes} in dots have not treated the
two-electron case.

Measurements were carried out on two similar lateral quantum dots formed
by Ti/Au depletion gates on the surface of a
$\mathrm{GaAs/Al_{0.3}Ga_{0.7}As}$ heterostructure $105$~nm above the 2DEG
layer (Fig.\:5, inset). The two devices showed similar results; most data
are from one of the dots, except those in Fig.\:5. The dilution
refrigerator base electron temperature was $T_{el}=45\pm5$~mK, measured
from Coulomb blockade peak widths. Differential conductance $g=dI/dV_{SD}$
was measured with typical ac excitations of $5\, \mu$V.

Figures 1(a,b) provide an overview of  transport spectroscopy data.
Diamond patterns of high conductance correspond to gate voltages $V_{G}$
where the ground state of the dot aligns with the chemical potential of
either the source or drain, allowing sequential tunneling through the dot
\cite{AleinerReview, DOS}. Transport is absent at more negative gate
voltages, indicating the absolute occupancy of the dot ($N= 0$ to $4$).
Conductance features that vanish below a finite source-drain voltage
$|V_{SD}|=\Delta/e$ involve transport through an excited state at energy
$\Delta$ above the ground state. An example of the latter is the nearly
horizontal band running through the center of the $N=2$ diamond. Beyond
this band transport through the excited triplet channel of the $N=2$ dot
becomes allowed, as discussed below.

\begin{figure}[t]
             \label{fig2}
         \includegraphics [width=3.2in]{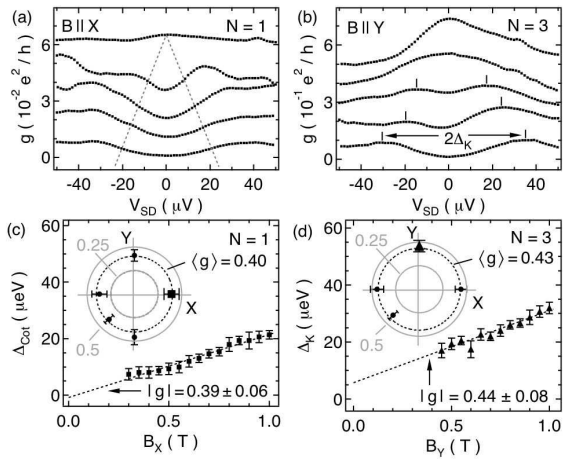}
      \caption{\footnotesize{(a) Differential conductance $g$ as a function of $V_{SD}$
      in the $N=1$ diamond ($V_G=0.1$ V) for in-plane fields
      $B_X=0,0.4,0.6,0.8,1$ T, (top to bottom, curves offset). Dashed grey lines are guides to the eye showing the cotunneling gap.
      (b) $g(V_{SD})$ shows a zero-bias
      peak in the $N=3$ valley ($V_G=0.42$ V) that splits in an in-plane field
      $B_Y=0,0.25,0.45,0.7,0.95$ T (top to bottom, curves offset).
      (c,d) splitting energies (see text) versus magnetic field
      as in (a,b) with linear fits.
      Insets: angular dependence of the g-factor in the plane of the 2DEG
      indicating isotropic behavior.
      Dashed circles show direction-averaged g-factors. Directions $X$ and $Y$ in the plane are arbitrary.
      }}
\vspace{-0.1in}
\end{figure}

Inside the diamonds, sequential tunneling is Coulomb blockaded and
transport requires higher order (cotunneling) processes \cite{Silvano,
AleinerReview}. Elastic cotunneling leaves the energy of the dot
unchanged; inelastic cotunneling, which leaves the dot in an excited state,
requires energy supplied by the source-drain bias. The inelastic
mechanism becomes active above a threshold $V_{SD}$ and is independent of
$V_G$. 

We first discuss the one-electron regime. A conduction threshold within
the $N=1$ diamond [Figs.\:1(a,b)] emerges from the crossing of
ground-state and excited-state sequential tunneling lines \cite{Silvano}.
These features correspond to the onset of inelastic cotunneling through
the first orbital excited state lying $\Delta_1 \sim 1.2(1.0)$~meV above
the ground state for a field $B_{\perp}=0(1)$~T perpendicular to the 2DEG.
Measurements with magnetic fields up to $1$~T along different directions
in the plane of the 2DEG show inelastic cotunneling through Zeeman split
one-electron states [Fig.\:2(a)]. Measurement of Zeeman energies via
cotunneling was established in Ref. \cite{KoganZeeman}. The cotunneling
gap $\Delta_{\mathrm{Cot}}$---extracted by taking half the peak splitting
of $dg/dV_{SD}$---is shown in Fig.\:2(c) for one of the field directions.
The g-factors are extracted from a linear fit to
$\Delta_{\mathrm{Cot}}(B)$ and are found to be isotropic within
experimental error, giving a value of $\langle g \rangle = 0.40\pm0.03$
averaged over the measured field directions. This is close to the bulk
GaAs value and consistent with previous (few-electron) experiments
\cite{Laterals}.

\begin{figure}[t]
             \label{fig3}
             \includegraphics[width=3.2in]{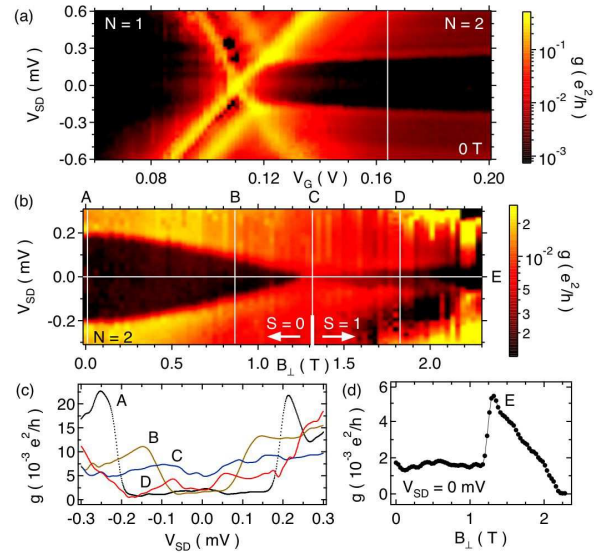}
     \caption{\footnotesize{
     (a) Differential conductance $g$ (log color scale) as a
     function of $V_{SD}$ and $V_G$ for $B=0$~T in the vicinity of the
     $N=1\rightarrow2$ transition.
     (b) $g(V_{SD},  B_\bot)$ at $V_G=0.164$~V (vertical white line in (a))
     \cite{diamag} shows the perpendicular field dependence of the singlet-triplet
     gap.  (c) Cuts showing $g(V_{SD})$ at the positions of the vertical lines in
     (b), marked A, B, C, D. (d) Horizontal cut (E in (b)) showing $g(B_\bot)$
     at zero bias. Note the asymmetric peak in $g$ at the singlet-triplet transition.
     }}
\vspace{-0.1in}
\end{figure}

For $N=3$, a zero-bias conductance peak, presumably due to the Kondo
effect \cite{Kondo}, splits in both perpendicular [Fig.\:1(d)] and
in-plane [Fig.\:2(b)] magnetic fields. The splitting $\Delta_{\mathrm{K}}$
due to in-plane field---taken as half the distance between maxima of the
split peaks [indicated in Fig.\:2(b)]---is shown in Fig.\:2(d) along with
a best fitting line. Slopes from the fits do not depend on direction in
the plane, and give $\langle g \rangle =0.43\pm0.03$, consistent with the
one-electron cotunneling data [Fig.\:2(c)]. Note that unlike the
cotunneling data, the Kondo data does not extrapolate to
$\Delta_{\mathrm{K}}(0)=0$, as also reported in previous experiments
\cite{KoganZeeman}. The threshold in-plane field $B_K$ for the appearance
of Kondo peak splitting gives an estimate of the Kondo temperature
($g\mu_B B_K \gtrsim k_B T_K$) of $T_K\sim150$~mK \cite{KoganZeeman}.

A detailed view of two-electron transport is shown in Fig.\:3(a). The
nearly horizontal band running through the $N=2$ diamond [see also
Fig.\:1(a,b)] corresponds to the onset of inelastic cotunneling through
the triplet excited state, which becomes active for $|V_{SD}|>J/e$. The
inelastic cotunneling edges align with the triplet excited state lines
seen in sequential tunneling outside the diamond, as expected
\cite{Silvano}. We use this cotunneling feature to measure the ST
splitting $J$. The zero-field value measured here, $J(B=0)\sim 0.2$ meV,
is much less than the $N=1$ orbital level spacing due to strong
interactions, consistent with theoretical estimates \cite{Burkard} and
previous measurements \cite{Sachrajda}. A surprising zero-bias conductance
peak in the middle of the cotunneling gap, visible in Fig.\:3(a) in the
range $0.12$~V$\lesssim V_G \lesssim 0.15$~V is not understood.

\begin{figure}
             \label{fig4}
         \includegraphics[width=3.1in]{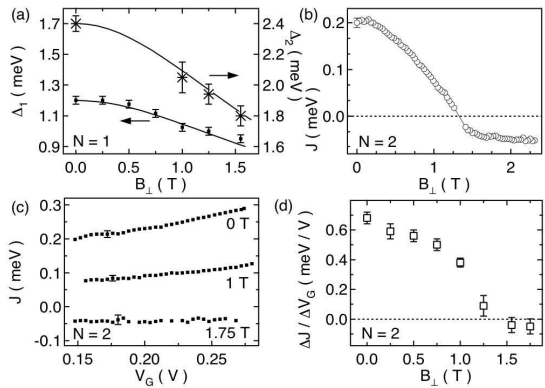}
      \caption{\footnotesize{(a) First and second one-electron excited state
      energies $\Delta_1$ and $\Delta_2$, measured from sequential tunneling
      along with fits to a 2D anisotropic harmonic oscillator model with
      $\hbar\omega_a=1.2$~meV and $\hbar\omega_b=\, 3.3$~meV (see text).
      (b) Singlet-triplet splitting $J$ as a function of magnetic field $B_\bot$.
      (c) Dependence of $J$ on gate voltage $V_G$ for various $B_\bot$ as indicated.
      (d) Average slopes $\Delta J(V_G)/ \Delta V_G $ from (c) as a
      function of magnetic field $B_\bot$, showing strong reduction of gate voltage
      dependence of J at large $B_\bot$.
      }}
\vspace{-0.1in}
\end{figure}

Perpendicular field dependence of the ST splitting $J(B_\perp)$ is
investigated by plotting $g$ along a cut through the $N=2$ valley as a
function of $B_\perp$ [Fig.\:3(b)]. Near $B_\perp = B^* \sim 1.3$ T the ST
gap closes and then re-opens at larger fields. We interpret this as a ST
crossing where the triplet state becomes
the ground state for $B_\perp>B^*$ [Fig.\:4(b)].
We note that in-plane fields up to $1$~T cause no observable change in the
two electron spectrum. We also find that $J$ depends on the gate voltage
$V_G$ [Fig.\:4(c)], as observed previously \cite{Kogan, Sachrajda}, though
at larger fields this dependence becomes significantly weaker
[Fig.\:4(d)]. The zero-bias conductance within the $N=2$ diamond as a
function of field shows a large, asymmetric peak at $B_\perp=B^*$ [Fig.\:
3(d)], consistent with predictions for elastic cotunneling at the ST
crossing \cite{Golovach} (see also \cite{STKondoSingleMode}).

Before turning to wave function correlations, we first extract some useful
information about the dot shape from the $N=1$ excitation spectrum.
Transport spectra for the $ N=0 \rightarrow 1$ transition, extracted from
plots like Fig.\:1(a) in the region between the $N=0$ and $N=1$ diamonds,
give first (second) excited state energies lying $\Delta_{1(2)}$ above the
ground state. We find $\Delta_{2} \sim 2 \Delta_{1}$, indicating roughly
harmonic confinement. Dependencies of $\Delta_{1(2)}$ on perpendicular
field are well described by a 2D anisotropic harmonic oscillator model
\cite{Schuh}. From zero-field data, we extract $\hbar \omega_a=1.2$~meV
where $a(b)$ is along the larger (smaller) dimension of the dot; the
energy scale for the smaller direction is found by fitting the field
dependence of $\Delta_{1}(B_\perp)$, which gives $\hbar \omega_b=3.3$~meV
\cite{Schuh}. As a check of these values, good agreement between
experimental and predicted values for $\Delta_{2}(B_\perp)$ is found
[Fig.\:4(a)]. We conclude that the dot potential is spatially elongated by
a factor of $\sim 1.6 = \sqrt{\omega_b/\omega_a}$.

We note that for strong coupling of the dot to the leads, the onset of
inelastic cotunneling at $V_{SD} = J/e$ shows considerable overshoot, as
seen in Fig.\:5 (measured in a device similar to the one discussed above,
with larger ST splitting, $J(0) \sim0.57$~meV). The temperature dependence
of the maximum overshoot is shown in the inset of Fig.\:5 along with a
line indicating a Kondo-inspired log$(T)$ dependence \cite{Sasaki,
STKondo, tubes}. The FWHM of the corresponding positive peak in
$dg/dV_{SD}$ is proportional to $T$ at high temperatures and saturates at
$T\sim80$~mK, giving an estimate of $T_K$ for this device. However, a
quantitative theory of nonequilibrium ST Kondo effect would be needed to
further analyze these data.

\begin{figure}[t]
             \label{fig5}
         \includegraphics[width=3.1in]{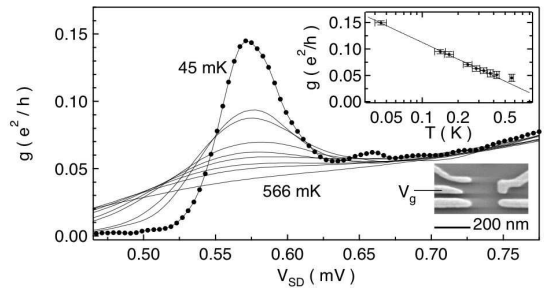}
      \caption{\footnotesize{ Differential conductance $g$ as a function
      of $V_{SD}$ for temperatures $T_{el}=45, 140, 170, 240, 280, 330,
      380, 420, 570$~mK (top to bottom) showing overshoot
      at $V_{SD}\sim J/e$. Inset: peak conductance as a function of temperature with best-fit $log(T)$ dependence (line).
      }}
\vspace{-0.1in}
\end{figure}

Finally, we investigate correlations in the two-electron wave function
following the analysis of Ref.\:\cite{Golovach}. We note that
Ref.\:\cite{Golovach} specifically considers a two-electron {\em double}
quantum dot; we anticipate that the elongated shape of our single dot will
lead to a spatially separated charge arrangement for $N=2$, not unlike a
double dot in the limit of strong interdot coupling. Selecting basis
states appropriate for a double dot but applicable here as well---i.e.,
symmetric ($|+\rangle$) and antisymmetric ($|-\rangle$) states along the
long axis of the dot---we identify $|+\rangle$ and $|-\rangle$ with the
orbital ground and first excited states of the one-electron dot. Because
of electron-electron interactions, the $N=2$ ground-state singlet
generally comprises an admixture of the one-electron ground and excited
orbital states. The amount of admixed excited state $|-\rangle$ is
parameterized by $\phi$ $(0 \leq \phi \leq 1)$, the so-called interaction
parameter. Knowing $\phi$ allows two other important quantities to be
extracted: the double occupancy, $D=(1-\phi)^2/2(1+\phi^2)$, and the
concurrence \cite{Schliemann}, $c=2\phi/(1+\phi^2)$, which respectively
parameterize correlations and entanglement of the two-electron singlet
ground state \cite{Golovach}.

To extract $\phi$ from elastic cotunneling data, one also needs to know
the charging energy for adding the second electron, the operating position
within the $N=2$ diamond, and the couplings to each lead for both the
singlet and the triplet,  $\Gamma_{1,2}^{S,T}$.  At fields well below the
ST transition, these $\Gamma$'s can be estimated from excitation spectra
at the $N = 1 \rightarrow 2$ transition by fitting a thermally broadened
Lorentzian to the tunneling lines \cite{Foxman}. Upon inserting these
quantities into Eqs.\:8 and 10 of Ref.\:\cite{Golovach}, we find
$\phi\sim0.5\pm0.1$, indicating that the $N=2$ ground-state singlet
contains a significant admixture of the excited one-electron orbital state
due to electron-electron interactions. We emphasize that this method does
not rely explicitly on a double dot interpretation \cite{LossComm}. From
this value of $\phi$ we extract a concurrence of $c\sim0.8$ for the
two-electron singlet. This is close to the maximum concurrence value
$c=1$, which characterizes a pair of singlet-correlated electrons in fully
non-overlapping orbital states.

Two alternative methods for estimating $\phi$ give consistent results with
the cotunneling method. First, one may adapt the formula
$\phi=\sqrt{1+(4t/U)^2}-4t/U$ from \cite{Golovach} by associating the
measured $\Delta_1$ with the tunnel splitting $2 t$ of the two lowest
noninteracting single-particle states, and the charging energy to add the
second electron with U. The second alternative method uses the size of the
elastic cotunneling step at the ST transition [see Fig.\:3(d)] which is
shown to be related to $\phi$ in \cite{Golovach}. It is notable that all
three methods allow the concurrence, a measure of ``useful'' (i.e.,
spatially separated) two-particle entanglement, to be extracted from a dc
transport measurement.

We thank Daniel Loss for numerous contributions to this work. We also thank M. Eto, L. Glazman, V. Golovach, W. Hofstetter, and M. Stopa for  valuable discussion. This work was supported in part by DARPA under the QuIST Program, the ARDA/ARO Quantum Computing
Program, and the Harvard NSEC. Work at UCSB was supported by QUEST, an NSF Science
and Technology Center.

{
small  }

\begin{thebibliography}{10}
\bibitem{DOS} Density of states modulations in the leads, here due to nanostructures
adjacent to the dot, can give additional weak diagonal lines (see for
example $V_{SD}>0$, $V_G\sim-0.05$ V).
\bibitem{diamag}
The gate voltage of the cut, $V_G^{\mathrm{cut}}(B)$, was adjusted to
compensate for a diamagnetic shift in gate voltage of the overall diamond
pattern. Gate voltages ranging from $V_G^{\mathrm{cut}}(0) = 0.164$ V to
$V_G^{\mathrm{cut}}(2\,$ T$) = 0.179$ V are determined by holding fixed
the position of the cut within the diamond pattern.
\end{thebibliography}
\end{document}